# A Global Map of Science Based on the ISI Subject Categories

Loet Leydesdorff [1] & Ismael Rafols [2]

**Abstract**

The ISI subject categories classify journals included in the *Science Citation Index (SCI)*. The aggregated journal-journal citation matrix contained in the *Journal Citation Reports* can be aggregated on the basis of these categories. This leads to an asymmetrical transaction matrix (citing versus cited) which is much more densely populated than the underlying matrix at the journal level. Exploratory factor analysis leads us to opt for a fourteen-factor solution. This solution can easily be interpreted as the disciplinary structure of science. The nested maps of science (corresponding to 14 factors, 172 categories, and 6,164 journals) are brought online at http://www.leydesdorff.net/map06/index.htm. An analysis of interdisciplinary relations is pursued at three levels of aggregation using the newly added ISI subject category of "Nanoscience & nanotechnology." The journal level provides the finer grained perspective. Errors in the attribution of journals to the ISI subject categories are averaged out so that the factor analysis can reveal the main structures. The mapping of science can, therefore, be comprehensive at the level of ISI subject categories.

## Introduction

The decomposition of the *Science Citation Index* into disciplinary and subdisciplinary structures has fascinated scientometricians and information analysts ever since the very beginning of this index. Price (1965) conjectured that the database would contain the very structure of science. He suggested that journals would be the appropriate units of analysis, and that aggregated citation relations among journals might reveal the disciplinary and finer-grained delineations such as those among specialties.

Carpenter & Narin (1973) tried to cluster the database in terms of aggregated journal citation patterns using the methods available at the time. However, the size of the database—2,200 journals in 1969 (Garfield, 1972: 472) and 6,164 journals in 2006—made it difficult to use algorithms more sophisticated than single-linkage clustering. Single-linkage clustering can operate on rank orders in lists, and thus one did not have to load these relatively large matrices into memory.

Small & Sweeney (1985) added a variable threshold to single-linkage clustering in their effort to map the sciences globally using co-citation analysis at the document level. However, the choice of thresholds, similarity criteria, and the clustering algorithms remained somewhat arbitrary. Because of the focus on relations, the latent dimensions of the matrix (its so-called "eigenvectors") could not be revealed using single-linkage clustering (Leydesdorff, 1987). A structural approach requires multivariate analysis, for example, based on rotation of the initial factor solution: units of analysis may occupy similar positions without necessarily maintaining strong relations (Burt, 1982; Leydesdorff, 2006).

---

[1] Amsterdam School of Communications Research (ASCoR), University of Amsterdam, Kloveniersburgwal 48, 1012 CX Amsterdam, The Netherlands; loet@leydesdorff.net; http://www.leydesdorff.net .

[2] SPRU (Science and Technology Policy Research), University of Sussex, Freeman Centre, Falmer Brighton, East Sussex BN1 9QE, United Kingdom; i.rafols@sussex.ac.uk .

The factor-analytical approach is limited even today to approximately 3,000 variables using the latest version of SPSS, while in the meantime the database has grown to more than 6,000 journals (Leydesdorff, 2006). Most researchers have therefore focused on chunks of the database or used seed journals for the collection of a sample (Doreian & Farraro, 1985; Tijssen *et al*., 1986; Leydesdorff, 1986). Boyack *et al*. (2005) used the *VxInsight* algorithm (Davidson *et al*., 1998) in order to map the whole journal structure as a representation of the structure of science.[3] However, the journal set remains fuzzy (Bensman, 2001) and therefore delineations are still based necessarily on trade-offs between accuracy, readability, and simplicity (Boyack *et al*., 2007). Klavans & Boyack (2007: 438) noted that a journal may occupy a different position in a different context. Leydesdorff (2007a) drew the conclusion that the point of entry remains crucial to the construction of a reference set, and therefore the database could be made accessible from any user's perspective with equal validity.

In addition to journals, the database has been mapped using co-citations (Small, 1973; Small & Griffith, 1974; Small & Sweeney, 1985) or co-occurrences of title words (Callon *et al*., 1983; 1986; Leydesdorff, 1989), at various levels of aggregation. However, using lower-level units (like documents) instead of journals means abandoning Price's grandiose vision to map the whole of science using the structure present in the aggregated journal-journal (co)citation matrix. The presence of a structure is indicated by the many zeros and missing values in this citation matrix. In 2006, for example, the database contained only 1,201,562 of the 37,994,896 (= $6{,}164^2$) possible relations. This corresponds to a density of 3.16%.

Since relations are dense in discipline-specific clusters and otherwise virtually non-existent, this matrix can be considered as nearly-decomposable. Near-decomposability is a general property of complex and evolving systems (Simon, 1973). The next-order units (e.g., disciplines or specialties) are reproduced in relatively stable sets which may change over time. The sets are functional subsystems that show a high density in terms of relations within the center, but are more open to change at the margins. (They can also become unstable.) The decomposition of such nearly-decomposable sets is analytically imprecise, but reasoned assumptions may nevertheless make it feasible (Newman, 2006a; 2006b).

## ISI Subject Categories

Hitherto, assumptions about components and clusters have been made mainly on formal grounds. How can one cluster the database bottom-up? However, the Institute of

[3] See at http://mapofscience.com for maps of science based on this algorithm.

Scientific Information (ISI) has added a substantive classifier to the database: the subject category or subject categories of each journal included. These categories are assigned by the ISI staff on the basis of a number of criteria including the journal's title, its citation patterns, etc. (McVeigh, *personal communication*, 9 March 2006).

The subject categories of the ISI cannot be considered as based on "literary warrant" like the classification of the Library of Congress (Chan, 1999). A classification scheme based on literary warrant is inductively developed in reference to the holdings of a particular library, or to what is or has been published. In other words, it is based on what the actual literature of the time warrants. For example, each of the individual schedules in the classification of the Library of Congress (LC) was initially drafted by subject specialists, who consulted bibliographies, treatises, comprehensive histories, and existing classification schemes to determine the scope and content of an individual class and its subclasses. The LC has a policy of continuous revision to take current literary warrant into account, so that new areas are developed and obsolete elements are removed or revised. The ISI categories are changed in terms of respective coverage, but cannot be revised from the perspective of hindsight.

In order to enhance flexibility in the database, the *Science Citation Index* is organized with a CD-Rom version for each year separately (which is by definition fixed at the date of delivery), and the *SCI-Expanded* version at the Internet to which relevant data can be added from the perspective of hindsight in order to optimize the database for information retrieval purposes. The *Journal Citation Reports*, however, are provided as a separate service of the ISI. The web version of this database is kept in complete agreement with the yearly CD-Rom. Thus, the subject categories themselves are not systematically updated, although new categories can be added and obsolete ones may no longer be used.

In addition to the subject categories, Thomson-ISI also assigns each journal in the *Essential Science Indicators* database (12,845 journals) to one of 22 so-called broad fields. (The listing of these attributions is available at http://www.in-cites.com/journal-list/index.html.) Journals are uniquely classified to a single broad field, while they can be classified under different subject categories in the *Science Citation Index*. The *Essential Science Indicators* provides statistics for government policy makers, university or corporate research administrators, etc., while the main service of the *Science Citation Index* is information retrieval for the research process.

For our research question about science mapping, the 172+ ISI subject categories are more interesting than the 22 broad fields because they are more finely grained. Leydesdorff (2006: 612) concluded that one cannot develop a conclusive classification on the basis of analytical arguments. However, the quality of a proposed classification can be tested against the structure in the data, in this case, aggregated journal-journal citations. Glänzel & Schubert (2003), for example, proposed twelve instead of 22 broad fields, but these categories are again different from the scheme of twelve categories proposed by Boyack *et al*. (2005). In this study, we focus on underlying data in the *Science Citation Index,* while these other studies included also the *Social Science Citation Index.* Given our analytical objective, the differences in citation behavior between the social sciences and the natural sciences could disturb the design as another source of variance.

The number of category attributions in the *Science Citation Index* is 9,848 for 6,164 journals in 2006 or, in other words, approximately 1.6 category per journal. The coverage of the 172 categories ranges from 262 journals sorted under "Biochemistry and Molecular Biology" to five journals which are sorted under a single category.[4] The average number of journals per category is 56.3.

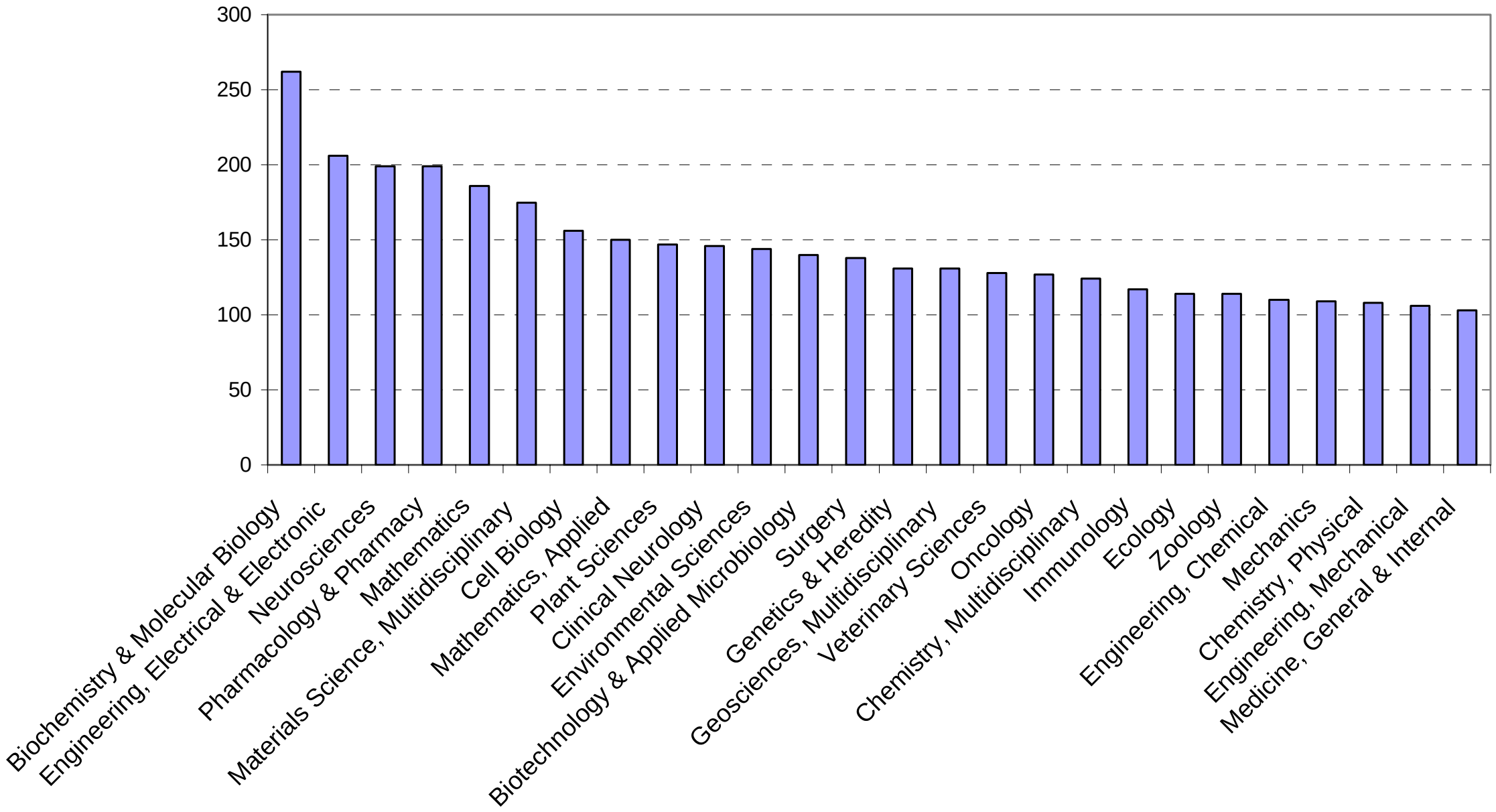

**Figure 1:** Frequency of 27 ISI Subject Categories which occur more than hundred times (*JCR* 2006).

The ISI subject categories match poorly with classifications derived from the database itself on the basis of an analysis of the principal components of the networks generated by citations (Leydesdorff, 2006: 611f.). Using a different methodology, Boyack *et al.* (2005) found that in somewhat more than 50% of the cases the ISI categories corresponded closely with the clusters based on inter-journal citation relations. These results accord with the expectation: many journals can be assigned unambiguous affiliations in one core set or another, but the remainder, which is also a large group, is heterogeneous (Bradford, 1934; Garfield, 1972).

This multidimensional set offers a wealth of options for generating representations. One should not expect a unique map of science, but a number of possible representations. Each map contains a projection from a specific perspective. However, one can ask whether there is a robust structure in terms of the latent dimensions of the underlying matrix. Using the ISI subject categories, we find that the system can be decomposed into fourteen factors or macro-disciplines. These macro-disciplines display three clusters: a biology-medicine cluster, a physics-materials-engineering-computing cluster, and a environment-ecology-geosciences cluster. This structure accords with the standard view of the organization of the sciences into disciplines.

[4] Three more categories which are no longer actively indexed subsume one or two journals.

## Methods

The data was harvested from the CD-Rom version of the *Journal Citation Reports* of the *Science Citation Index* 2006. As indicated above, 175 subject categories are used. Three categories ("Psychology, biological," "Psychology, experimental," and "Transportation") are no longer used as classifiers in the citing dimension, but four journals are still indicated with these three categories in the cited dimension. Thus, we work with 172 citing categories and 175 cited.

The matrix, accordingly, contains two structures: a cited and a citing one. Salton's cosine was used for normalization in both the cited and the citing directions (Salton & McGill, 1983; Ahlgren *et al*., 2003). Pajek is used for the visualizations (Batagelj and Mvar, 2007) and SPSS (v15) for the factor analysis. The threshold for the visualizations is pragmatically set at cosine > 0.2. Visualizations are based on the algorithm of Kamada & Kawai (1989). The size of the nodes is set proportional to the number of citations in a given category (or in Figure 3, the logarithm of this value). The thickness and grey-shade of the links is proportional to the cosine values in five equal steps of 0.2.

Using a factor model, the crucial question is the number of factors to be extracted. Unless one has *a priori* reasons for testing an assumption, this number has to be determined on empirical grounds (Leydesdorff, 2006). SPSS includes by default all factors with an eigenvalue larger than unity. The screeplot of the eigenvalues can be used for the assessment of the number of meaningful factors. However, the assumption has to be tested against the data.

## Results

Unlike the aggregated journal-journal citation matrix, the matrix of 172 (citing) times 175 (cited) categories is not sparse: 11,577 of the (172 x 175 =) 30,100 cells have a zero value. This corresponds with 38.46% of the number of cells.[5] Since the categories are unevenly distributed, one cannot set a threshold value across the matrix without normalization. The factor analysis, however, begins with a normalization using the Pearson correlation coefficient. As noted, the visualizations are based on cosine values. The cosine is equal to the Pearson correlation, but without normalization to the arithmetic mean (Jones & Furnas, 1987).

Let us focus on the structure in the citing dimension because this structure is actively maintained by the indexing service and is therefore current. The screeplot of the eigenvalues suggests a fourteen-factor solution (Figure 2).

---

[5] UCINet computes a density for this matrix after binarization of 0.7538 ± 0.4308.

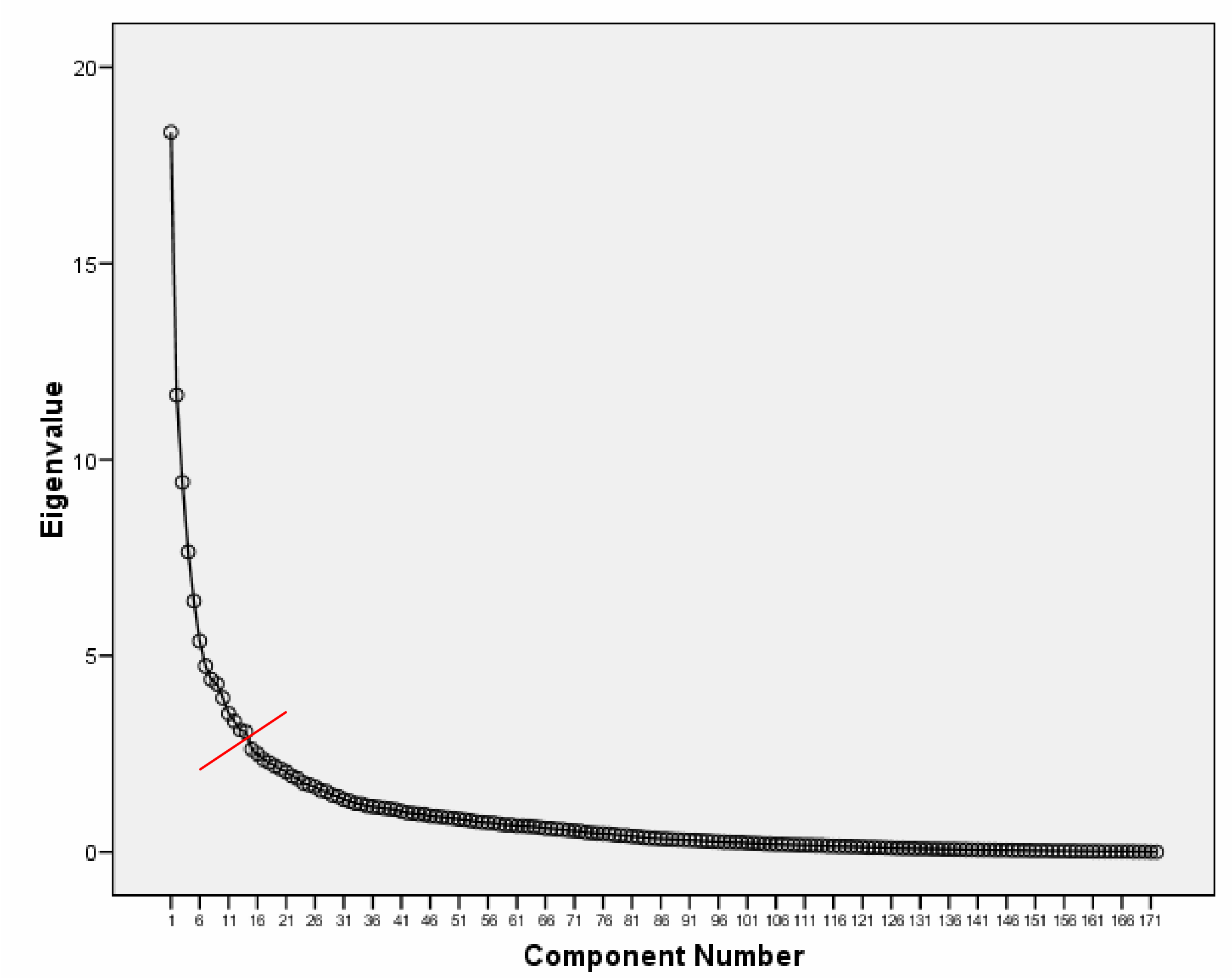

**Figure 2**. Scree plot of the Factor Analysis (Citing)

Table 1 shows the four highest loadings on the last factor in the case of extracting 13, 14, or 15 factors, respectively. This confirms that the quality of the factors declines considerably after extracting 14 factors. The fourteen-factor solution explains 51.8% of the variance of the matrix in the citing projection, and 47.9% of the variance in the cited projection.

| *Highest factor loadings on factor 13 in the case of a 13-factor solution* | *Highest factor loadings on factor 14 in the case of a 14-factor solution* | *Highest factor loadings on factor 15 in the case of a 15-factor solution* |
|---|---|---|
| 0.779 | 0.786 | 0.593 |
| 0.715 | 0.721 | 0.539 |
| 0.672 | 0.698 | 0.484 |
| 0.669 | 0.687 | 0.472 |

**Table 1**: Highest factor loadings on the last factor in a 13-, 14-, and 15-factor solution, respectively.

The factor loadings for the 172 categories on the fourteen factors in the citing dimension are provided in Appendix I. They can be interpreted in terms of disciplines, such as Physics, Chemistry, Clinical Medicine, Neurosciences, Engineering, Ecology, etc.

The factors in the cited dimension can be designated using precisely the same disciplinary classification, but their rank-order (that is, the percentage of variance explained by each factor) is different (Table 2). 154 out of the 172 categories (that is, 89%) fall in the same factor in both the citing and cited projections. The eighteen categories that are classified differently in the citing and cited projections are listed in Table 3.

| | **Citing factors** | **Cited factors** |
|---|---|---|
| Biomedical Sciences | 1 | 1 |
| Materials Sciences | 2 | 2 |
| Computer Sciences | 3 | 4 |
| Clinical medicine | 4 | 5 |
| Neuro Sciences | 5 | 3 |
| Ecology | 6 | 7 |
| Chemistry | 7 | 9 |
| Geosciences | 8 | 6 |
| Engineering | 9 | 8 |
| Infectious diseases | 10 | 10 |
| Environmental sciences | 11 | 12 |
| Agriculture | 12 | 11 |
| Physics | 13 | 13 |
| General medicine; health | 14 | 14 |

**Table 2**: Fourteen disciplines distinguished on the basis of ISI Subject Categories in 2006 ($\rho >$ 0.95; $p < 0.01$).

| **ISI Subject Category** | **Citing Factor** | **Cited Factor** |
|---|---|---|
| Urology & nephrology | Biomedical Sci. | Clinical Medicine |
| Pharmacology & pharmacy | Biomedical Sci. | Neuro-sciences |
| Physiology | Biomedical Sci. | Neuro-sciences |
| Medicine, legal | Biomedical Sci. | Chemistry |
| Toxicology | Biomedical Sci. | Environmental Sci |
| Biotechnology & applied microbiology | Biomedical Sci. | Agriculture |
| Nutrition & dietetics | Biomedical Sci. | Agriculture |
| Mathematical & computational biology | Biomedical Sci. | General Medicine; Health |
| Energy & fuels | Materials Sci. | Engineering |
| Computer sci., interdiscipl. Applic. | Computer Sci. | Engineering |
| Mathematics | Computer Sci. | Engineering |
| Engineering, industrial | Computer Sci. | Physics |
| Chemistry, physical | Chemistry | Materials Sci. |
| Materials science, biomaterials | Chemistry | Materials Sci. |
| Chemistry, applied | Chemistry | Agriculture |
| Materials science, composites | Engineering | Materials Sci. |
| Mycology | Infect. Diseases | Agriculture |
| Medicine, general & internal | Medicine General | Clinical Medicine |

**Table 3**: Eighteen ISI Subject Categories that are classified differently in the citing and cited dimensions.

Figure 3 shows the map of 171 ISI subject categories which relate above the threshold of cosine > 0.2. The category “Agricultural economics and politics” is not related at this level. The nodes represent the categories and are colored in terms of the fourteen factors. (The picture in the cited dimension is very similar.) In this chart, the node sizes were set proportional to the *logarithm* of the number of citations (in the respective subject category) in order to keep the visualization readable.

**Figure 3**: 171 ISI subject categories in the citing dimension; cosine > 0.2. Node sizes set proportional to the logarithm of the number of citations given by each category.

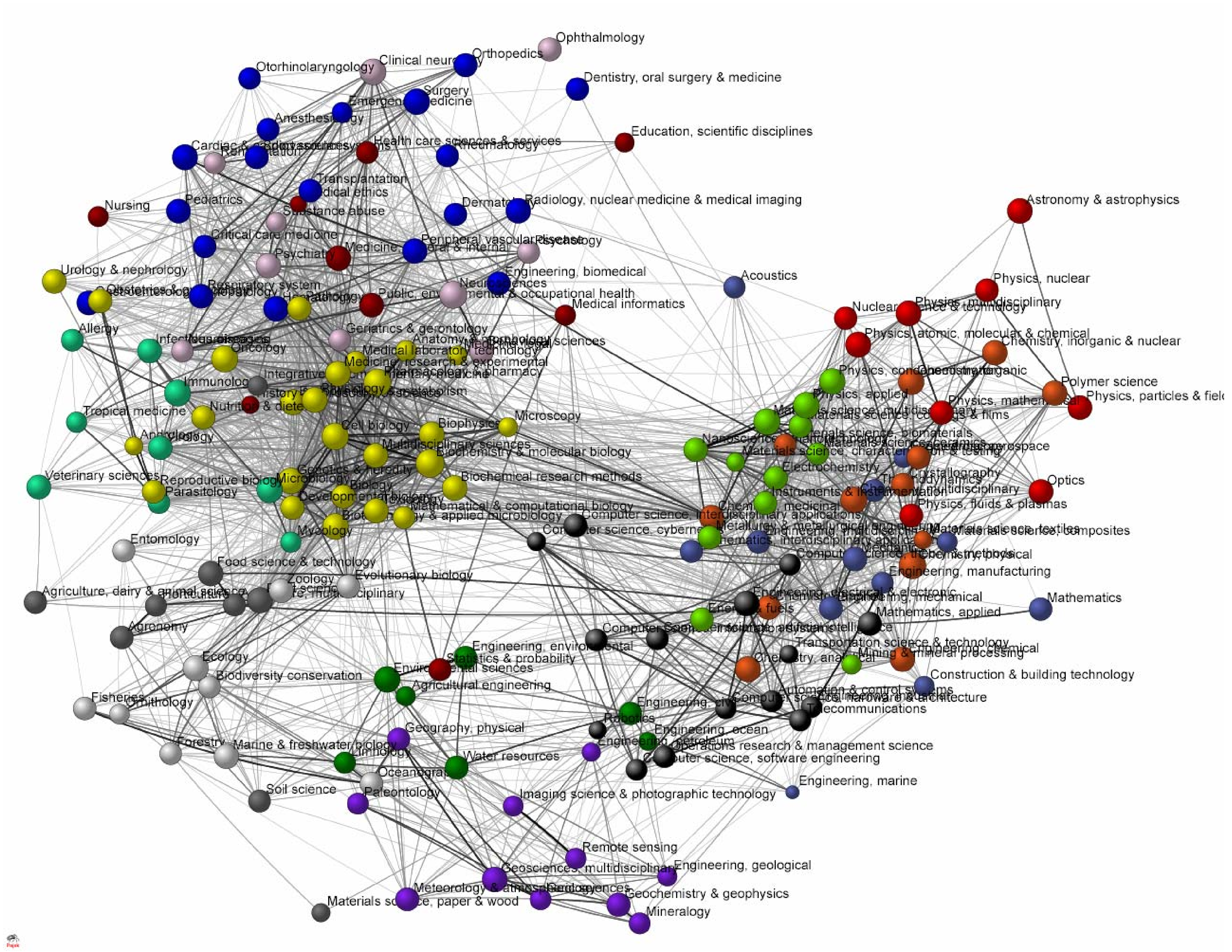

Whereas the traditional disciplines are represented by clear factors (e.g., Physics or Chemistry), specific fields of application in mathematics or engineering do not fall in the disciplinary classification, but in the factor representing their topic. For example, Mathematical Physics is classified as Physics, while Mathematics is classified as part of Engineering and with a second factor loading on the Computer Sciences. Chemical Engineering loads on the Chemistry factor more than on the one representing Engineering.

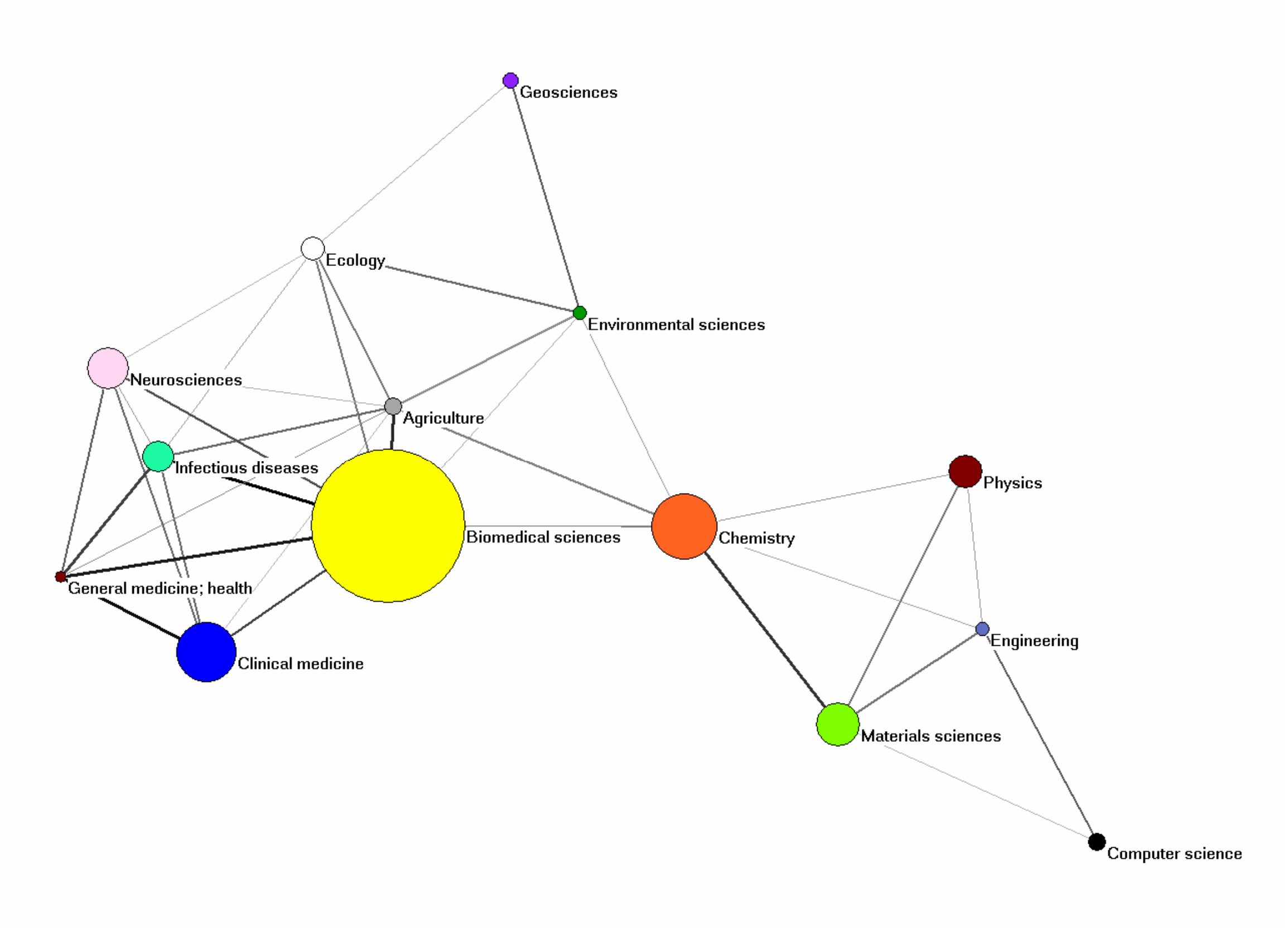

**Figure 4**: Fourteen disciplines in the citing dimension; cosine > 0.2. (The colors correspond with those used in Figure 3.)

Figure 4 shows the citation relations among the fourteen groups. (The depiction in the cited dimension is again virtually similar to this one in the citing dimension.) While Figure 3 informed us in greater detail about the relations among subdisciplines and specialties, the factor-analytical categories allow us to depict these ISI subject categories in Figure 3 with different colors in terms of the disciplinary affiliations provided in Figure 4. Both levels are interactively related with hyperlinks at http://www.leydesdorff.net/map06/index.htm .

The largest factor is designated as Biomedical Sciences. It includes at the disaggregated level:

i. The core biological sciences, such as Biochemistry & Molecular Biology, Developmental Biology, Genetics, and Cell Biology;
ii. The methodologies crucial for the biological sciences, such as Microscopy and Biochemical Research Methods;

iii. Disciplines that fall into medicine but are highly interrelated with the biological sciences, such as Oncology and Pathology.

Among the latter, eight subject categories (e.g., Physiology, Toxicology, or Nutrition Sciences) have a citing pattern in the factor of the Biomedical Sciences (that is, they draw on basic biological knowledge), but they show a cited pattern in factors more related to specific applications such as Neurosciences, Environmental Sciences, or Infectious Diseases (see Table 3 above).

Four factors are closely associated with the Biomedical Sciences: Clinical Medicine, Neurosciences, Infectious Diseases, and General Medicine & Health. At the opposite pole of the medicine-related factors, we find factors based on the hard sciences: the factors of Physics, Engineering, Materials Sciences, and Computer Sciences, are among them. Chemistry plays a brokerage role between Physics and Material Sciences, on the one side, and core Biomedical Sciences such as Biophysics and Biochemistry, on the other.

The relative positions of the subject categories within Figure 3 inform us *prima facie* about their disciplinary or interdisciplinary character. However, one should be cautious in drawing conclusions from a visual inspection of maps. A map remains a two-dimensional projection of a space (in this case, a fourteen-dimensional one), and one therefore needs a large number of projections from different angles before one can formulate hypotheses on this basis. In a next section, we will use betweenness centrality as a measure of interdisciplinarity for comparing the results at the level of ISI subject categories with the lower level results of aggregated journal-journal citations.

On the basis of a number of these projections—that is, variants of Figure 3—we feel comfortable in suggesting that the connection between the “medical pole” and the “hard-science pole” is achieved by the way of three main routes:

i. A direct link between the Computer Sciences and some of the medical specialties such as Psychology, Neuro-imaging, and Medical Informatics;
ii. Through Chemistry, which appears to play a brokerage role between Physics and Material Sciences, on the one side, and the core Biomedical Sciences such as Biophysics and Biochemistry, on the other;
iii. Through a path that links Engineering and Material Sciences with Geosciences and Environmental Sciences, and also connects these two latter factors with Ecology and Agriculture. The latter are related to Infectious Diseases and the large set of journals in the Biomedical Sciences. This path can be considered as a small cluster with a focus on environmental issues.

Our results are consistent with previously reported maps (Boyack *et al*. 2005; Boyack and Klavans, 2007), but we chose to exclude the social sciences. We would expect differences and similarities when mapping the social sciences (using the *Social Science Citation Index*) because of the different order of magnitude of citations in the journal-journal citation network, differences in citation behavior, and the different functions of citations as relations among texts in these sciences.

## Interdisciplinarity

Various studies of interdisciplinarity have been based on the assumption that journals can be grouped using the ISI subject categories (e.g., Van Leeuwen & Tijssen, 2000; Morillo *et al.*, 2001, 2003). "Interdisciplinarity" is often a policy objective, while new developments may take place at the borders of disciplines (Zitt, 2005). One of the potential uses of a map of science is to help us understand the cognitive base and the relative positions of emerging fields (Van Raan, 2000; Bordons *et al.*, 2004; Porter *et al.*, 2006 and 2007).

In another context, one of us argued for using betweenness centrality in the vector space as a measure of interdisciplinarity in *local* citation environments (Leydesdorff, 2007b). The citation matrix among ISI subject categories provides us with a *global* map. Could one use this same measure for interdisciplinarity—that is, betweenness centrality after transformation of the data into a vector space using the cosine—as a measure of interdisciplinarity at the level of the ISI subject categories? A case study of an interdisciplinary development relevant to our data can teach us more about the relations between the three levels of aggregation: (1) aggregated journal-journal citation data, (2) citation relations among ISI subject categories, and (3) citation relations among the fourteen factors.

"Nanoscience & nanotechnology" can be considered as an interesting case in point: within the policy arena this research domain was constructed as a critical and highly interdisciplinary field (Kostoff, 2004; 2007), but this interdisciplinarity has remained controversial in the various mapping efforts (Schummer, 2004; Leydesdorff & Zhou, 2006; Rafols & Meyer, 2007). The ISI added the subject category of "Nanoscience & nanotechnology" (NS) in 2005. Twenty-seven journals were subsumed under this category in 2005 and 32 in 2006.

Note that the focus in Leydesdorff's (2007b) proposal to use betweenness centrality as a measure of interdisciplinarity was formulated in terms of citation impact and not in terms of citation behavior. Thus, we will use in this analysis the other side of the citation matrix among subject categories. Among the 175 subject categories in this ("cited") direction, the newly added category NS contributes only 1.16% to the betweenness centrality. (Among the 172 subject categories in the citing dimension this is 1.5%.),

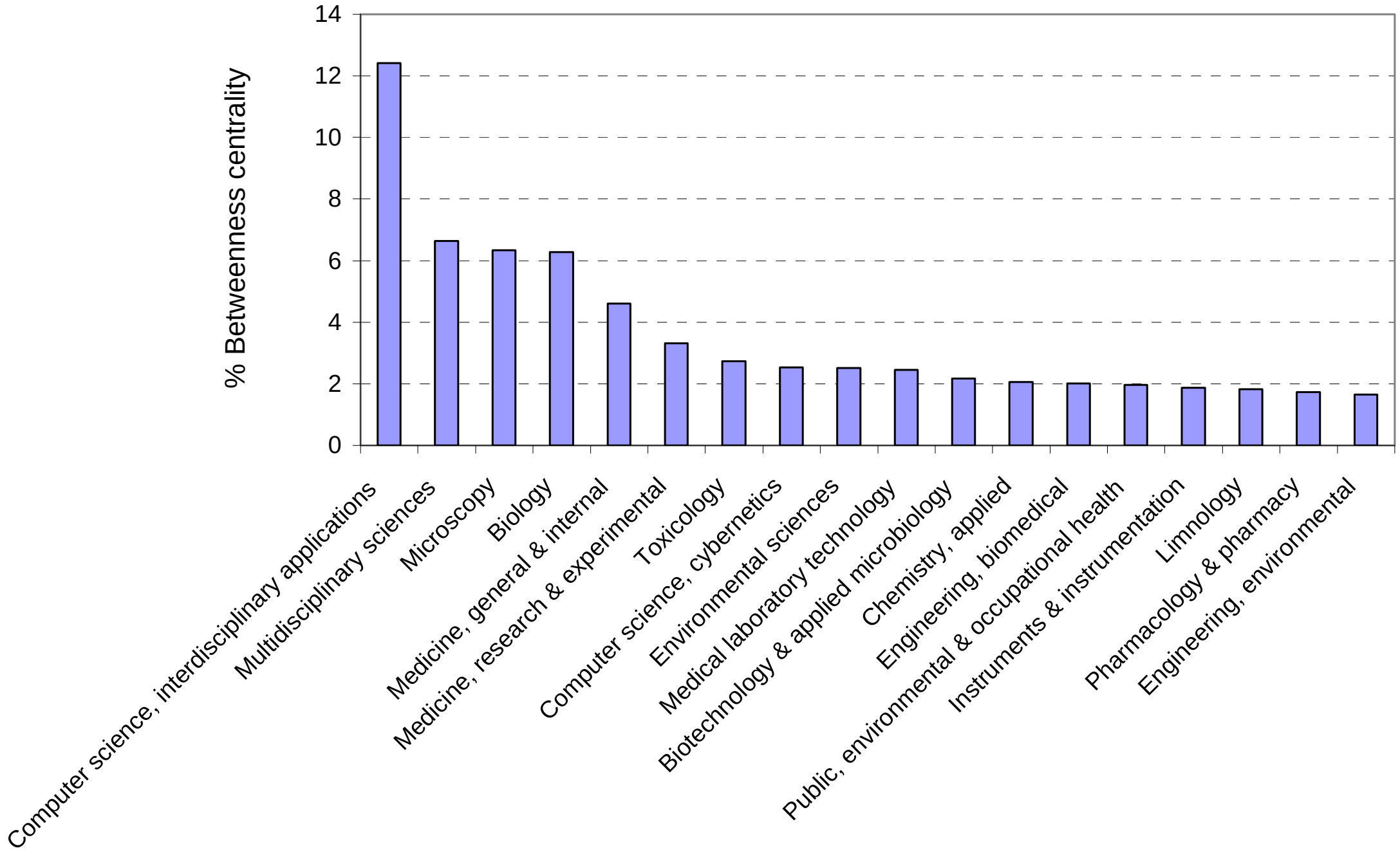

**Figure 5**: Eighteen ISI Subject Categories with contribute 1.5% or more to the betweenness centrality in the vector space in the cited dimension.

Figure 5 shows the 18 ISI subject categories that contribute with 1.5% percent or more to the betweenness centrality in the matrix. Not surprisingly, the category of "Computer science, interdisciplinary applications" leads the ranking. This corresponds with visual inspection of Figure 3. Categories like "Microscopy" (6.34%) and "Instruments & instrumentation" (1.87%) are also included, but the latter not to the extent one would expect on the basis of theoretical contributions (Price, 1984; Shinn & Joerges, 2002). The position of "Biology" in this ranking is surprising. Can the discipline of biology also be considered as an interdisciplinary field when analyzed in terms of subject categories?

As against betweenness centrality among journals in local citation environments, this betweenness centrality among categories informs us about which categories are holding the science system together more than others at the level of the database. The computer sciences and general bio-medical categories seem to play a key role. At the level of subject categories we note a shift from chemistry, with high betweenness centrality at the level of disciplines (Figure 4 above), towards biology at the level of subject categories.

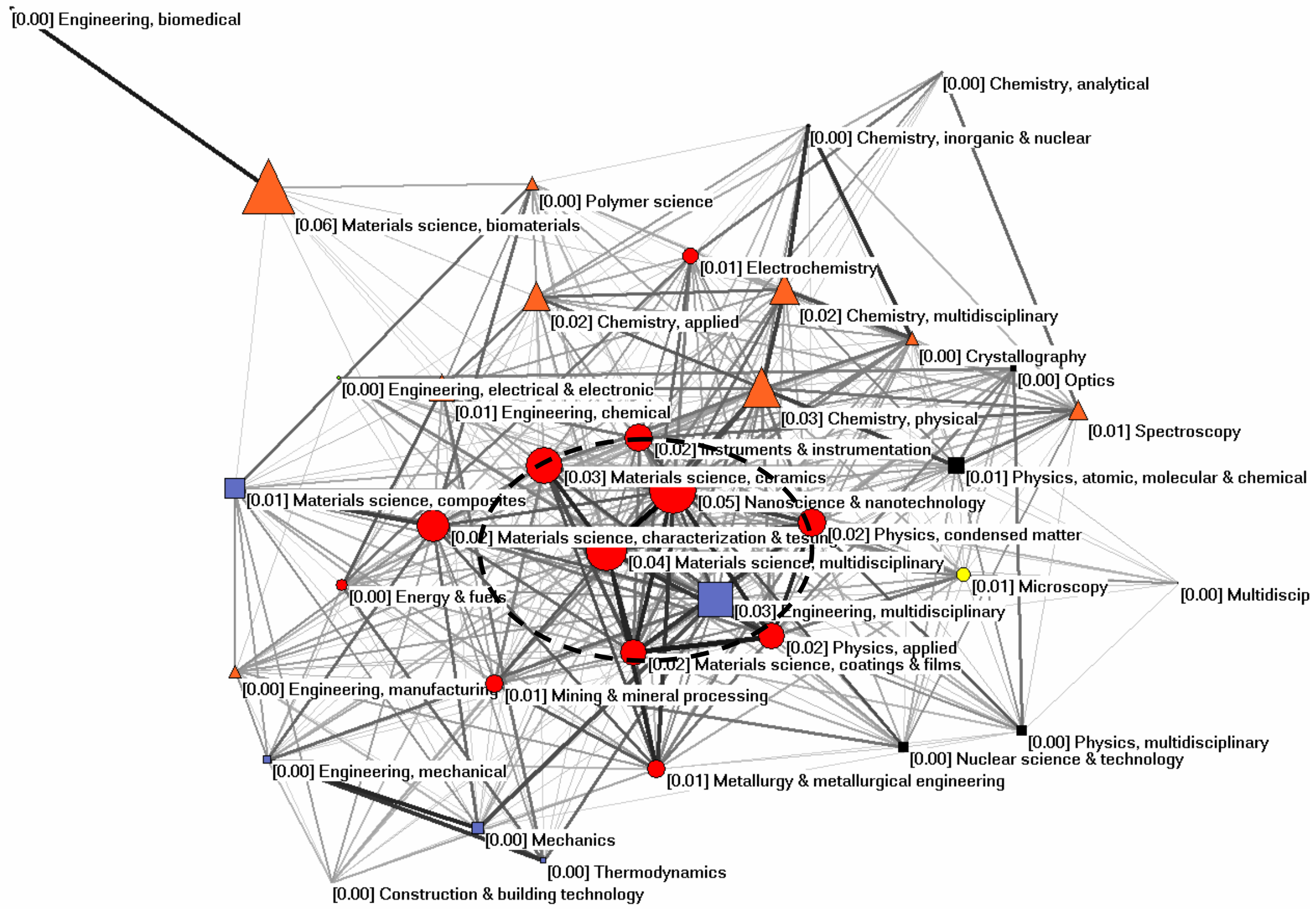

**Figure 6**: 37 ISI Subject Categories in the cited environment of "Nanoscience & nanotechnology" (cosine > 0.2). Legend of factors: Red circles = Materials Sciences; Brown triangles = Chemistry; Light squares = Engineering; Black Squares = Physics; Light circle = Biomedical Sciences.

The disciplinary and interdisciplinary structures are nested. Within each local area—represented by a subset—one can expect again a division between journals that form the core of a specific set and are in this sense more central to the discipline, and journals that reach with citations across the borders between different sets. Figure 6 provides the $k = 1$ neighbourhood of the ISI subject category "Nanoscience & nanotechnology" as a selection of 37 of the 175 subject categories related at the level of cosine > 0.2. Among them, twelve subject categories have their highest factor loading on the factor that was designated as "Materials Sciences" in Appendix I. However, the match is not perfect. "Chemistry, physical" for example, contributes considerably to the betweenness centrality (3%), but is classified as Chemistry; "Engineering, multidisciplinary" (3%) similarly contributes to the betweenness centrality more than "Metallurgy & metallurgic engineering," which is classified as "Materials Sciences."

These results can be appreciated as indicating problems of classification in a fuzzy set. "Nanoscience and nanotechnology" itself contributes 5.4% to the betweenness centrality in this environment, but "Materials science, biomaterials" leads the ranking with 5.6% although it is poorly connected to the core set. Thus, this map is comprehensible but not conclusive. The categories are on average more closely related to one another in terms of aggregate citations than the journals. However, the cosine-normalized map does not exhibit a structure as we know it from the journal mapping, to which we will now proceed.

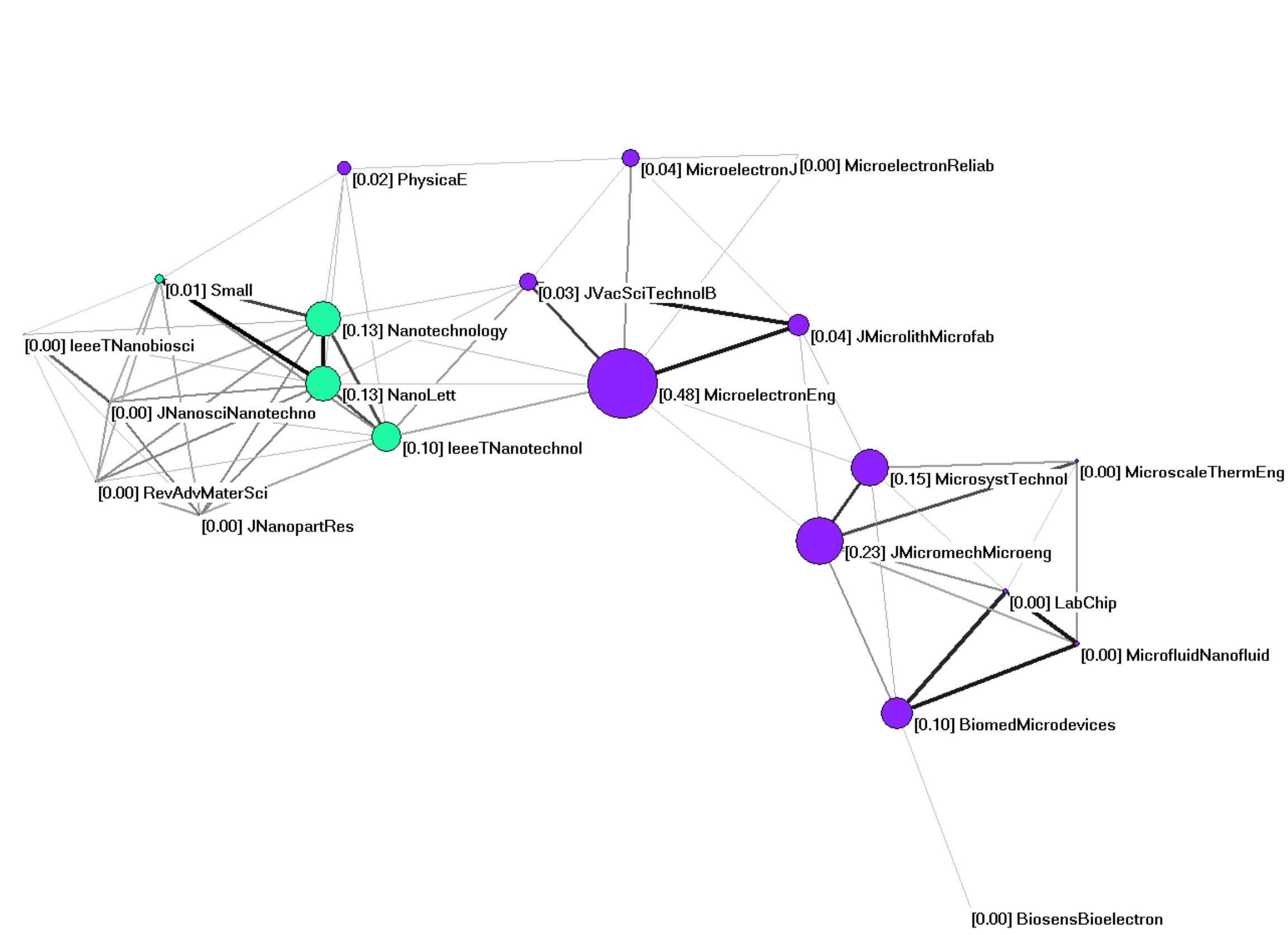

**Figure 7**: Betweenness centrality in the citation impact environments of 21 journals classified as "Nanoscience and nanotechnology" in 2006; cosine > 0.2.

Figure 7 shows the relations among the citation patterns of 21 journals classified as "Nanoscience and nanotechnology." Of the 32 journals classified under this category by the ISI, the journal *Nanoscale and Microscale Thermophysical Engineering* was cited only one once in 2006 and was therefore left aside. Ten other journals are not related to the central network at the threshold level of cosine > 0.2. For example, *Fullerenes Nanotubes and Carbon Nanostructures* is not cited in a manner similar to these journals, but it is related in terms of its citing pattern. However, the density of the underlying citation matrix among these 32 journals is 34.2%.

The transition to the vector space—that is, the normalization in terms of cosine values—clarifies the structure in the matrix: *Microelectronic Engineering* is the single journal with the highest betweenness centrality (47.6%). However, it is not part of the cluster of journals with "nano" in their title. The latter form a separate cluster, but they are not leading in terms of betweenness centrality.

In Leydesdorff (2007b: 1317), the journal *Nano Letters* was selected as the seed journal for drawing a relevant impact environment into the analysis. This journal was chosen because it had by far the highest impact factor in this environment in 2004 (8.449), and this analysis was based on citation impact. By 2006, the impact factor of *Nano Letters* had further increased to 9.960, but other journals like *Small* and *Lab on a Chip* now also have impact factors around 6.0. If we repeat the analysis of 2004 for 2006, 18 instead of 17 journals are relevant in the citation environment of *Nano Letters* to the extent that they contribute more than 1% to its total citations. Figure 8 provides the visualization in terms of betweenness centrality in the vector space.

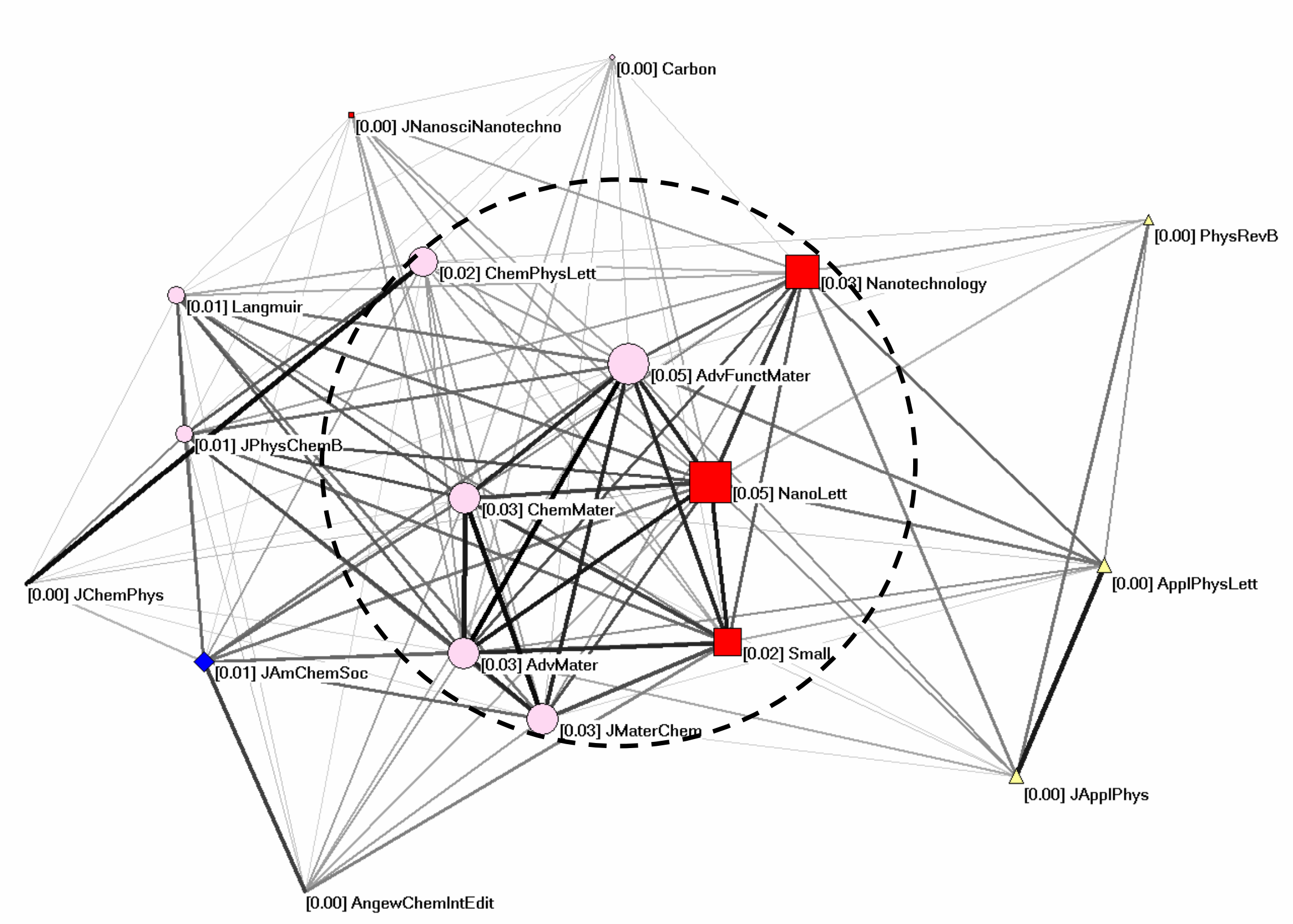

**Figure 8**: Betweenness centrality of 18 journals in the citation impact environment of *Nano Letters* 2006 (cosine > 0.2). Colors and shapes of nodes correspond to the ISI Subject Categories: squares for “Nanoscience & nanotechnology,” circles for “Chemistry, physical”; triangles for “Physics, applied,” and diamonds for “Chemistry, multidisciplinary.”

Figure 8 shows an interdisciplinary set, but the set is composed of journals assigned to the new subject category for “nano,” on the one hand, and “physical chemistry,” on the other. The orientation of *Nano Letters* to the emerging set of nano-journals in 2004 is now overshadowed by its intermediating function between chemistry journals and nano-journals. (*Nano Letters* is published by the American Chemical Society!) In other words, the structure of the field has changed in the local citation impact environment of *Nano Letters*. This can be signaled by using the betweenness centrality indicator at the level of journals, but not by using the ISI subject categories.

## Conclusions and discussion

The analysis of "Nanoscience & nanotechnology" at various levels of aggregation (journals, ISI subject categories, and factors representing disciplines) suggests that the ISI subject categories function moderately well for the classification. They do not match with the fine-grained picture that can be constructed when using aggregated journal-journal citations, but the latter have the disadvantage that they are difficult to map at higher levels of aggregation (like specialties and disciplines in a bottom-up mode) without making assumptions.

One may wonder why the ISI subject categories that were found to be a poor match for journal citation patterns in other research (Boyack *et al*., 2005; Leydesdorff, 2006) perform relatively well when used as aggregates, and can then further be aggregated in order to provide comprehensive maps of science in both the cited and citing dimensions. The explanation is statistical: Boyack *et al*. (2005) noted that the ISI subject categories match in approximately 50% of the cases and mismatch consequently in the remaining 50%. The error, however, is not systematic so that the 50% matching cases prevail in the aggregate. Factor analysis enables us to distinguish the pattern as a signal from the noise. Thus, a clear factor structure can be discerned at this intermediate level.

From the top-down perspective of the factor structure, the noise at the bottom level can be considered as mere variation which is distributed stochastically. Factor analysis enables us to reduce the complexity in the data. As we noted above, the resulting maps match well with the previously published mappings of the team of Boyack, Börner, and Klavans. The maps of the latter can be found at http://mapofscience.com. The matrix of aggregated inter-category citations is available at http://www.leydesdorff.net/map06/data.xls. This file enables users to draw one's own maps or make one's own extractions and inferences. The maps are available as a nested structure at http://www.leydesdorff.net/map06/index.htm.

## References:

Ahlgren, P., Jarneving, B., & Rousseau, R. (2003). Requirement for a Cocitation Similarity Measure, with Special Reference to Pearson's Correlation Coefficient. *Journal of the American Society for Information Science and Technology,* 54(6), 550-560.

Batagelj, V., & Mrvar, A. (2007) *Pajek. Program for Large Network Analysis.* Retrieved on October 4, 2007 from, http://vlado.fmf.uni-lj.si/pub/networks/pajek/.

Bensman, S. J. (2001). Bradford's law and fuzzy sets: Statistical implications for library analyses. *IFLA Journal,* 27, 238-246.

Bordons, M., Morillo, F., & Gómez, I. (2004). Analysis of cross-disciplinary research through bibliometric tools. In H. F. Moed, W. Glänzel & U. Schmoch (Eds.), *Handbook of quantitative science and technology research* (pp. 437-456). Dordrecht: Kluwer.

Boyack, K. W., Klavans, R., & Börner, K. (2005). Mapping the Backbone of Science. *Scientometrics,* 64(3), 351-374.

Boyack, K., Börner, K., & Klavans, R. (2007). Mapping the Structure and Evolution of Chemistry Research. In D. Torres-Salinas & H. Moed (Eds.), *Proceedings of the 11th International Conference of Scientometrics and Informetrics,* Vol. 1, pp. 112-123, CSIC, Madrid, 21-25 June 2007.

Boyack, K., & Klavans, R. (2007) A map of science. Retrieved on October 4, 2007, from http://mapofscience.com/.

Bradford, S. C. (1934). Sources of information on specific subjects. *Engineering,* 137, 85-86.

Burt, R. S. (1982). *Toward a Structural Theory of Action*. New York, etc.: Academic Press.

Callon, M., Courtial, J.-P., Turner, W. A., & Bauin, S. (1983). From Translations to Problematic Networks: An Introduction to Co-word Analysis, *Social Science Information* 22, 191-235.

Callon, M., Law, J., & Rip, A. (Eds.). (1986). *Mapping the Dynamics of Science and Technology*. London: Macmillan.

Carpenter, M. P., & Narin, F. (1973). Clustering of Scientific Journals. *Journal of the American Society for Information Science,* 24, 425-436.

Chan, L. M. (1999). *A Guide to the Library of Congress Classification. 5th ed.* . Englewood, CO: Libraries Unlimited.

Davidson, G. S., Hendrickson, B., Johnson, D. K., Meyers, C. E., & Wylie, B. N. (1998). Knowledge Mining With VxInsight: Discovery Through Interaction. *Journal of Intelligent Information Systems,* 11(3), 259-285.

Doreian, P., & Fararo, T. J. (1985). Structural Equivalence in a Journal Network. *Journal of the American Society for Information Science,* 36, 28-37.

Garfield, E. (1972). Citation Analysis as a Tool in Journal Evaluation. *Science* 178(Number 4060), 471-479.

Jones, W. P., & Furnas, G. W. (1987). Pictures of Relevance: A Geometric Analysis of Similarity Measures. *Journal of the American Society for Information Science,* 36(6), 420-442.

Kamada, T., & Kawai, S. (1989). An algorithm for drawing general undirected graphs. *Information Processing Letters,* 31(1), 7-15.

Klavans, R., & Boyack, K. (2007). Is there a Convergent Structure of Science? A Comparison of Maps using the ISI and Scopus Databases. In D. Torres-Salinas & H. Moed (Eds.), *Proceedings of the 11th International Conference of Scientometrics and Informetrics,* Vol. 1, pp 437-448, CSIC, Madrid, 21-25 June 2007.

Kostoff, R. (2004). The (Scientific) Wealth of Nations. *The Scientist,* 18(18), 10.

Kostoff, R. N., Koytcheff, R. G., & Lau, C. G. Y. (2007). Global nanotechnology research metrics. *Scientometrics,* 70(3), 565-601.

Leydesdorff, L. (1986). The Development of Frames of References. *Scientometrics* 9, 103-125.

Leydesdorff, L. (1987). Various methods for the Mapping of Science. *Scientometrics* 11, 291-320.

Leydesdorff, L. (1989). Words and Co-Words as Indicators of Intellectual Organization. *Research Policy,* 18, 209-223.
Leydesdorff, L. (2004). Clusters and Maps of Science Journals Based on Bi-connected Graphs in the Journal Citation Reports. *Journal of Documentation,* 60(4), 371-427.
Leydesdorff, L. (2006). Can Scientific Journals be Classified in Terms of Aggregated Journal-Journal Citation Relations using the Journal Citation Reports? *Journal of the American Society for Information Science & Technology,* 57(5), 601-613.
Leydesdorff, L. (2007a). Visualization of the Citation Impact Environments of Scientific Journals: An online mapping exercise. *Journal of the American Society of Information Science and Technology,* 58(1), 207-222.
Leydesdorff, L. (2007b). "Betweenness Centrality" as an Indicator of the "Interdisciplinarity" of Scientific Journals, *Journal of the American Society for Information Science and Technology* 58(9), 1303-1309.
Leydesdorff, L. and Zhou, P. (2007). Nanotechnology as a Field of Science: Its Delineation in Terms of Journals and Patents. *Scientometrics,* 70(3), 693-713.
Morillo, F., Bordons, M., & Gómez, I. (2001). An approach to interdisciplinarity through bibliometric indicators. *Scientometrics,* 51(1), 203-222.
Morillo, F., Bordons, M., & Gómez, I. (2003). Interdisciplinarity in Science: A Tentative Typology of Disciplines and Research Areas. *Journal of the American Society for Information Science and Technology,* 54(13), 1237-1249.
Newman, M. E. J. (2006a). Finding community structure in networks using the eigenvectors of matrices. *Physical Review E,* 74(3), 36104.
Newman, M. E. J. (2006b). Modularity and community structure in networks. *Proceedings of the National Academy of Sciences,* 103(23), 8577-8582.
Porter, A. L., Roessner, J. D., Cohen, A. S., & Perreault, M. (2006). Interdisciplinary research: meaning, metrics and nurture. *Research Evaluation,* 15(3), 187-195.
Porter, A. L., Cohen, A. S., David Roessner, J., & Perreault, M. (2007). Measuring researcher interdisciplinarity. *Scientometrics,* 72(1), 117-147.
Price, D. J. de Solla (1965). Networks of scientific papers. *Science,* 149, 510- 515.
Price, D.J. de Solla (1984). The science/technology relationship, the craft of experimental science, and policy for the improvement of high technology innovation. *Research Policy* 13(1), 3-20.
Rafols, I. and Meyer, M. (2007) How cross-disciplinary is bionanotechnology? Explorations in the specialty of molecular motors. *Scientometrics* 70(3), 633-650.
Salton, G., & McGill, M. J. (1983). *Introduction to Modern Information Retrieval*. Auckland, etc.: McGraw-Hill.
Schummer, J. (2004) Multidisciplinarity, Interdisciplinarity, and patterns of research collaboration in nanoscience and nanotechnology. *Scientometrics* 59, 425-465.
Shinn, T., & Joerges, B. (2002) The transverse science and technology culture: dynamics and roles of research-technologies. *Social Science Information* 41(2), 207-251.
Simon, H. A. (1973). The Organization of Complex Systems. In H. H. Pattee (Ed.), *Hierarchy Theory: The Challenge of Complex Systems* (pp. 1-27). New York: George Braziller Inc.
Small, H. (1973). Co-citation in the Scientific Literature: A New measure of the Relationship between Two Documents. *Journal of the American Society for Information Science,* 24(4), 265-269.

Small, H., & Griffith, B. (1974). The Structure of Scientific Literature I. *Science Studies* 4, 17-40.
Small, H., & Sweeney, E. (1985). Clustering the Science Citation Index Using Co-Citations I. A Comparison of Methods,. *Scientometrics 7*, 391-409.
Tijssen, R., de Leeuw, J., & van Raan, A. F. J. (1987). Quasi-Correspondence Analysis on Square Scientometric Transaction Matrices. *Scientometrics* 11, 347-361.
Van Leeuwen, T., & Tijssen, R. (2000). Interdisciplinary dynamics of modern science: analysis of cross-disciplinary citation flows. *Research Evaluation,* 9(3), 183-187.
Van Raan, A. F. J. (2000). The Interdisciplinary Nature of Science. Theoretical Framework and Bibliometric-Empirical Approach. In P. Weingart & N. Stehr (Eds.), *Practicing Interdisciplinarity* (pp. 66-78). Toronto: University of Toronto Press.
Zitt, M. (2005). Facing Diversity of Science: A Challenge for Bibliometric Indicators. *Measurement: Interdisciplinary Research and Perspectives,* 3(1), 38-49.

## Appendix I

Factor analysis of 172 ISI Subject Categories over de the Citing projection

| ← CITING FACTOR | CITING FACTORS → ISI Subject Category | Biomedical Sciences | Materials Sciences | Computer Sciences | Clinical Sciences | Neuro-Sciences | Ecology | Chemistry | Geo-sciences | Engineering | Infectious Diseases | Environmental Sciences | Agriculture | Physics | Gen. Medicine; Health |
|---|---|---|---|---|---|---|---|---|---|---|---|---|---|---|---|
| BIOMEDICAL SCIENCES | cell biology | **0.890** | | | 0.118 | | | | | | 0.192 | | | | |
| | biochemistry & molecular biology | **0.857** | | | | | | 0.212 | | | 0.215 | | 0.170 | | |
| | biophysics | **0.818** | | | | | | 0.270 | | | 0.171 | | 0.136 | | |
| | developmental biology | **0.818** | | | | | | | | | | | | | |
| | multidisciplinary sciences | **0.802** | | | | 0.185 | 0.209 | 0.145 | 0.167 | | 0.243 | | 0.128 | 0.126 | |
| | genetics & heredity | **0.773** | | | | | 0.254 | | | | 0.149 | | 0.222 | | |
| | biology | **0.746** | | | | 0.241 | 0.439 | | | | 0.104 | | 0.107 | | |
| | medicine, research & experimental | **0.738** | | | 0.351 | 0.113 | | | | | 0.435 | | | | 0.113 |
| | microscopy | **0.718** | 0.283 | | | | | | | | | | | | -0.114 |
| | anatomy & morphology | **0.711** | | | 0.134 | 0.261 | 0.127 | | | | | | | | -0.143 |
| | endocrinology & metabolism | **0.645** | | | 0.151 | 0.162 | | | | | | | | | 0.166 |
| | biotechnology & applied microbiology | **0.622** | | | | | | 0.204 | | | 0.367 | | 0.376 | | |
| | physiology | **0.601** | | | 0.254 | 0.432 | | | | | | | | | |
| | reproductive biology | **0.594** | | | -0.115 | | | -0.199 | -0.110 | -0.142 | -0.145 | | -0.116 | | 0.147 |
| | andrology | **0.592** | | -0.103 | | | -0.105 | -0.187 | -0.114 | -0.153 | -0.187 | 0.128 | -0.124 | | 0.205 |
| | medical laboratory technology | **0.586** | | | 0.412 | | | | | | 0.199 | | | | 0.241 |
| | biochemical research methods | **0.575** | | | | | | 0.382 | | | 0.187 | | 0.204 | | |
| | pathology | **0.560** | | | 0.369 | 0.127 | | | | | 0.204 | | | | |
| | oncology | **0.559** | | | 0.248 | | | | | | | | | | 0.151 |
| | mathematical & computational biology | **0.543** | -0.162 | 0.173 | | | 0.254 | 0.229 | | 0.166 | | -0.189 | | | 0.221 |
| | toxicology | **0.511** | | | | 0.142 | | 0.182 | | | 0.173 | 0.339 | 0.112 | | 0.175 |
| | **pharmacology & pharmacy** | **0.500** | | | 0.163 | 0.429 | | 0.251 | | | 0.275 | | 0.109 | | 0.149 |
| | **obstetrics & gynecology** | **0.400** | | | | | -0.130 | -0.214 | -0.107 | -0.130 | -0.199 | 0.119 | -0.138 | | 0.319 |
| | **nutrition & dietetics** | **0.369** | | | 0.109 | | -0.121 | | | | | | 0.236 | | 0.258 |
| | **medicine, legal** | **0.346** | | | | 0.117 | | 0.164 | | | 0.149 | | | | 0.180 |
| | **urology & nephrology** | **0.241** | | | 0.241 | | | | | | | | | | 0.152 |
| **MATERIAL SCIENCES** | **materials science, multidisciplinary** | | **0.913** | | | | | 0.239 | | 0.161 | | | | | |
| | **nanoscience & nanotechnology** | | **0.878** | | | | | 0.301 | | 0.114 | | | | | |
| | **materials science, coatings & films** | | **0.860** | | | | | 0.107 | | | | | | | |
| | **physics, applied** | | **0.828** | 0.151 | | | | | | | | | | 0.250 | |
| | **materials science, ceramics** | | **0.785** | | | | | 0.133 | | | | | | | |
| | **metallurgy & metallurgical engineering** | | **0.771** | | | | | | | 0.154 | | | | | |
| | **physics, condensed matter** | | **0.721** | | | | | 0.136 | | | | | | 0.371 | |
| | **materials science, characterization & testing** | | **0.676** | | | | | | | 0.483 | | | | | |
| | **mining & mineral processing** | | **0.557** | | | | | | 0.273 | 0.112 | | | | | |
| | **instruments & instrumentation** | | **0.518** | 0.295 | | | | 0.102 | | | | | | 0.354 | |
| | **electrochemistry** | | **0.467** | | | | | 0.295 | | | | | | | |
| | **energy & fuels** | | **0.301** | | | | | 0.234 | | 0.278 | | 0.239 | | | |
| **COMPUTER SCIENCES** | **computer science, hardware & architecture** | | | **0.882** | | | | | | | | | | | |
| | **computer science, information systems** | | | **0.814** | | | | | | | | | | | |
| | **computer science, artificial intelligence** | | | **0.812** | | | | | | | | | | | |
| | **engineering, electrical & electronic** | | 0.335 | **0.793** | | | | | | | | | | 0.123 | |
| | **computer science, theory & methods** | | | **0.791** | | | | | | | | | | | |
| | **computer science, software engineering** | | | **0.750** | | | | | | | | | | | |
| | **telecommunications** | | 0.114 | **0.744** | | | | | | -0.109 | | 0.104 | | | |
| | **computer science, cybernetics** | | | **0.729** | | 0.330 | | | | | | | | | |
| | **automation & control systems** | | | **0.718** | | | | | | | | | | | |
| | **transportation science & technology** | | | **0.553** | | | | -0.110 | | 0.189 | | 0.118 | | | |
| | **computer science, interdisciplinary applications** | 0.271 | -0.133 | **0.514** | | | | 0.317 | | 0.439 | | | | 0.112 | 0.113 |
| | **robotics** | | | **0.484** | | | | | | | | | | | |
| | **operations research & management science** | | | **0.420** | | | | | | 0.201 | | -0.154 | | -0.178 | |
| | **mathematics, applied** | | -0.159 | **0.329** | | | | | | 0.291 | | -0.159 | | 0.193 | |
| | **engineering, industrial** | | | **0.314** | | | | | | 0.256 | | -0.148 | | -0.248 | |
| **CLINICAL MEDICINE** | **surgery** | | | | **0.796** | | | | | | | | | | |
| | **critical care medicine** | | | | **0.706** | 0.123 | | | | | 0.131 | | | | 0.148 |
| | **emergency medicine** | | | | **0.666** | | | | | | | | | | 0.283 |
| | **transplantation** | 0.118 | | | **0.636** | | | | | | 0.304 | | | | |
| | **respiratory system** | 0.120 | | | **0.617** | | | | | | 0.167 | | | | 0.126 |
| | **peripheral vascular disease** | 0.285 | | | **0.616** | | | | | | | | | | 0.173 |
| | **cardiac & cardiovascular systems** | 0.146 | | | **0.608** | | | | | | | | | | 0.151 |
| | **orthopedics** | | | | **0.551** | 0.160 | | | | | | | | | |
| | **engineering, biomedical** | 0.196 | | | **0.546** | | | 0.146 | | | -0.143 | | | | -0.116 |
| | **hematology** | 0.461 | | | **0.490** | | | | | | 0.183 | | | | 0.132 |
| | **pediatrics** | | | | **0.445** | 0.174 | | -0.103 | | | 0.112 | | | | 0.293 |
| | **sport sciences** | | | | **0.381** | 0.271 | | | | | -0.117 | | | | |
| | **anesthesiology** | | | | **0.374** | 0.295 | | | | | | | | | |
| | **gastroenterology & hepatology** | 0.137 | | | **0.343** | | | | | | 0.166 | | | | |
| | **radiology, nuclear medicine & medical imaging** | 0.120 | | | **0.335** | 0.257 | | | | | -0.150 | | | | |
| | **otorhinolaryngology** | | | | **0.328** | 0.161 | | | | | | | | | |
| | **rheumatology** | 0.120 | | | **0.217** | | | | | | 0.163 | | | | 0.141 |
| | **dermatology** | 0.178 | | | **0.207** | | | | | | 0.205 | | | | |
| | **dentistry, oral surgery & medicine** | | | | **0.197** | | | | | | | | | | |
| **NEURO-SCIENCES** | **neurosciences** | 0.271 | | | | **0.870** | | | | | | | | | |
| | **psychology** | | | | | **0.807** | | | | | | | | | |
| | **behavioral sciences** | 0.164 | | | | **0.804** | 0.255 | | | | | | | | |
| | **neuroimaging** | | | | 0.198 | **0.792** | | | | | -0.104 | | | | |
| | **psychiatry** | | | | | **0.783** | | | | | | | | | 0.174 |
| | **clinical neurology** | | | | 0.310 | **0.752** | | | | | | | | | |
| | **substance abuse** | | | | | **0.641** | | | | | 0.142 | | | | 0.222 |
| | **geriatrics & gerontology** | 0.395 | | | 0.194 | **0.617** | | | | | | | | | 0.274 |
| | **rehabilitation** | | | | 0.379 | **0.516** | | | | | -0.106 | | | | |
| | **ophthalmology** | 0.116 | | | | **0.157** | | | | | | | | | |
| **ECOLOGY** | **ecology** | | | | | | **0.902** | | | | | | 0.141 | | |
| | **biodiversity conservation** | | | | | | **0.859** | | | | | 0.113 | 0.135 | | |
| | **zoology** | 0.259 | | | | 0.243 | **0.707** | | | | | | | | |
| | **marine & freshwater biology** | | | | | | **0.701** | | | | | 0.288 | | | |
| | **ornithology** | | | | | | **0.678** | | | | | | | | |
| | **evolutionary biology** | 0.410 | | | | | **0.635** | | | | | -0.145 | 0.250 | | |
| | **oceanography** | | | | | | **0.566** | | 0.287 | | | 0.297 | -0.134 | | |
| | **fisheries** | | | | | | **0.525** | | | | 0.111 | 0.197 | -0.130 | | |
| | **forestry** | | | | | | **0.517** | | | | | | 0.375 | | |
| | **entomology** | 0.194 | | | | | **0.366** | | | | | | 0.156 | | |
| **CHEMISTRY** | **chemistry, multidisciplinary** | 0.138 | 0.251 | | | | | **0.842** | | | | | | | |
| | **chemistry, organic** | | | | | | | **0.715** | | | | | | | |
| | **chemistry, inorganic & nuclear** | | 0.120 | | | | | **0.701** | | | | | -0.115 | | |
| | **chemistry, physical** | | 0.530 | | | | | **0.622** | | | | | | 0.138 | |
| | **chemistry, applied** | 0.150 | 0.132 | | | | -0.120 | **0.600** | | | | 0.141 | 0.480 | | |
| | **crystallography** | | 0.338 | | | | | **0.600** | | | | | | | |
| | **chemistry, medicinal** | 0.471 | | | | 0.164 | | **0.511** | | | 0.184 | | 0.191 | | |
| | **spectroscopy** | 0.128 | 0.202 | | | | | **0.496** | | -0.113 | | | | 0.322 | |
| | **engineering, chemical** | | 0.240 | | | | | **0.439** | | 0.268 | | 0.270 | 0.104 | | |
| | **chemistry, analytical** | 0.160 | | | | | | **0.413** | | | | 0.158 | 0.127 | | |
| | **materials science, textiles** | | 0.165 | | | | | **0.410** | | | | 0.111 | 0.111 | -0.149 | |
| | **polymer science** | | 0.289 | | | | | **0.372** | | 0.136 | | | | -0.124 | |
| | **materials science, biomaterials** | 0.272 | | | 0.282 | | | **0.323** | | | | | | | -0.156 |
| **GEOSCIENCES** | **geosciences, multidisciplinary** | | | | | | | | **0.920** | | | 0.207 | | | |
| | **geology** | | | | | | | | **0.908** | | | | | | |
| | **geochemistry & geophysics** | | | | | | | | **0.807** | | | | | | |
| | **geography, physical** | | | | | | 0.335 | | **0.786** | | | 0.156 | | | |
| | **paleontology** | | | | | | 0.166 | | **0.700** | | | | | | |
| | **mineralogy** | | 0.137 | | | | | | **0.649** | | | | | | |
| | **engineering, geological** | | | | | | | | **0.632** | 0.129 | | 0.105 | | | |
| | **engineering, petroleum** | | 0.104 | | | | | 0.203 | **0.555** | 0.115 | | 0.153 | | | |
| | **remote sensing** | | | 0.231 | | | | | **0.471** | -0.119 | | 0.273 | | | |
| | **meteorology & atmospheric sciences** | | | | | | | | **0.463** | | | 0.365 | | | |
| | **imaging science & photographic technology** | | | 0.280 | | | | | **0.354** | -0.101 | | 0.285 | | | |
| **ENGINEERING** | **mechanics** | | 0.181 | | | | | | | **0.831** | | | | 0.142 | |
| | **engineering, mechanical** | | 0.238 | | | | | | | **0.757** | | 0.107 | | | |
| | **mathematics, interdisciplinary applications** | | | 0.162 | | | | 0.136 | | **0.652** | | | | 0.301 | |
| | **thermodynamics** | | 0.147 | | | | | 0.113 | | **0.649** | | 0.114 | | | |
| | **engineering, multidisciplinary** | | 0.482 | 0.152 | | | | 0.153 | | **0.635** | | | | 0.107 | |
| | **engineering, aerospace** | | | 0.116 | | | | | | **0.453** | | | | 0.109 | |
| | **materials science, composites** | | 0.421 | | | | | | | **0.444** | | | | -0.189 | |
| | **engineering, marine** | | | | | | | -0.104 | | **0.387** | | 0.262 | -0.119 | | |
| | **construction & building technology** | | 0.245 | | | | | | | **0.339** | | 0.215 | | -0.156 | |
| | **engineering, manufacturing** | | 0.248 | 0.292 | | | | | | **0.323** | | -0.140 | | -0.250 | |
| | **acoustics** | | | 0.119 | | | | -0.106 | | **0.297** | -0.128 | | | | |
| | **mathematics** | | **-0.133** | **0.117** | | | | | | **0.127** | | **-0.125** | | **0.102** | |
| **INFECTIOUS DISEASES** | **infectious diseases** | | | | 0.126 | | | | | | **0.811** | | | | 0.158 |
| | **immunology** | 0.315 | | | 0.237 | | | | | | **0.749** | | | | |
| | **microbiology** | 0.356 | | | | | | | | | **0.652** | | 0.216 | | |
| | **allergy** | | | | 0.220 | | | | | | **0.582** | | | | |
| | **virology** | 0.371 | | | | | | | | | **0.573** | | | | |
| | **tropical medicine** | | | | | | | | | | **0.565** | | | | 0.366 |
| | **parasitology** | 0.300 | | | | | 0.111 | | | | **0.560** | | | | |
| | **mycology** | 0.419 | | | | | 0.127 | | | | **0.461** | | 0.432 | | |
| | **veterinary sciences** | 0.151 | | | | | | | | -0.103 | **0.429** | | | | |
| **ENVIRONMENTAL SCIENCES** | **engineering, environmental** | | | | | | 0.136 | 0.176 | 0.114 | | | **0.762** | | | |
| | **environmental sciences** | | | | | | 0.340 | 0.113 | 0.179 | | | **0.753** | 0.128 | | |
| | **water resources** | | | | | | 0.133 | | 0.345 | 0.122 | | **0.730** | | | |
| | **engineering, civil** | | | | | | | | 0.205 | 0.398 | | **0.682** | | | |
| | **limnology** | | | | | | 0.541 | | 0.251 | | | **0.610** | | | |
| | **agricultural engineering** | | | | | | | 0.109 | | | | **0.562** | 0.395 | | |
| | **engineering, ocean** | | | | | | | -0.112 | 0.298 | 0.365 | | **0.383** | -0.115 | | |
| **AGRICULTURE** | **horticulture** | 0.119 | | | | | 0.101 | | | | | | **0.802** | | |
| | **agronomy** | | | | | | 0.172 | | | | | | **0.791** | | |
| | **agriculture, multidisciplinary** | 0.130 | | | | | | 0.150 | | | | 0.154 | **0.757** | | |
| | **plant sciences** | 0.293 | | | | | 0.223 | | | | | | **0.714** | | |
| | **food science & technology** | 0.146 | | | | | -0.151 | 0.177 | -0.104 | | | 0.136 | **0.571** | | |
| | **soil science** | | | | | | 0.221 | | 0.159 | | | 0.263 | **0.452** | | |
| | **integrative & complementary medicine** | 0.199 | | | 0.138 | 0.321 | | 0.162 | | | 0.135 | | **0.322** | | 0.258 |
| | **agriculture, dairy & animal science** | 0.169 | | | -0.122 | | | | | -0.102 | | | **0.203** | | |
| | **materials science, paper & wood** | | | | | | | | | | | | **0.109** | | |
| **PHYSICS** | **physics, multidisciplinary** | | 0.207 | | | | | | | 0.114 | | | | **0.864** | |
| | **physics, mathematical** | | | | | | | | | 0.274 | | | | **0.781** | |
| | **physics, nuclear** | | | | | | | | | | | | | **0.724** | |
| | **physics, particles & fields** | | | | | | | | | | | | | **0.687** | |
| | **physics, fluids & plasmas** | | | | | | | | | 0.456 | | | | **0.599** | |
| | **optics** | | 0.337 | 0.260 | | | | | | | | | | **0.484** | |
| | **physics, atomic, molecular & chemical** | | 0.282 | | | | | 0.400 | | | | | | **0.440** | |
| | **astronomy & astrophysics** | | | | | | | | | | | | | **0.379** | |
| | **nuclear science & technology** | | 0.362 | | | | | | | | | 0.107 | | **0.373** | |
| **GENRAL MEDICINE; HEALTH** | **health care sciences & services** | | | | 0.183 | 0.104 | | | | | | | | | **0.786** |
| | **medical ethics** | | | | 0.139 | | | | | | | | | | **0.721** |
| | **public, environmental & occupational health** | | | | | 0.113 | | | | | 0.227 | 0.178 | | | **0.698** |
| | **medicine, general & internal** | 0.144 | | | 0.498 | 0.129 | | | | | 0.203 | | | | **0.687** |
| | **medical informatics** | | -0.164 | 0.226 | 0.100 | | | | | | | -0.157 | | | **0.564** |
| | **nursing** | | | | | | | | | | | | | | **0.461** |
| | **history & philosophy of science** | | | | | | 0.127 | | | | | | | 0.150 | **0.434** |
| | **education, scientific disciplines** | | | | | | | 0.168 | | | | | | | **0.411** |
| | **statistics & probability** | 0.116 | -0.184 | 0.212 | | | 0.115 | 0.150 | | 0.195 | | -0.178 | | | **0.241** |

Extraction Method: Principal Component Analysis.
Rotation Method: Varimax with Kaiser Normalization. Rotation converged in 11 iterations.